\documentclass[twocolumn,letter]{jpsj3}
\usepackage{txfonts}
\usepackage{amsmath,amssymb,amsfonts,float}

\newcommand{\bolk}{\mathbf{k}}
\newcommand{\bolp}{\mathbf{p}}
\newcommand{\bolq}{\mathbf{q}}
\newcommand{\bole}{\mathbf{e}}
\newcommand{\bolQ}{\mathbf{Q}}

\newcommand{\boll}{\mathbf{l}}

\newcommand{\boli}{\mathbf{i}}
\newcommand{\bolj}{\mathbf{j}}

\newcommand{\bolK}{\mathbf{K}}

\newcommand{\VEV}[1]{\langle #1 \rangle}  

\title{Magnetic phase diagram slightly below the saturation field
in the stacked $J_1$-$J_2$ model in the square lattice with the $J_{\text{C}}$ 
interlayer coupling}
\author{Hiroaki T. \name{Ueda}}
\inst{Okinawa Institute of Science and Technology, Onna-son, Okinawa 904-0412, 
Japan} 

\abst{
We study the effect of adding interlayer coupling to the square lattice, $J_1$-$J_2$ Heisenberg model in high external magnetic field.
In particular, we consider a cubic lattice formed from stacked $J_1$-$J_2$ layers, with interlayer exchange coupling $J_{\text{C}}$. 
For the 2-dimensional model ($J_{\text{C}}=0$) it has been shown that a spin-nematic phase appears close to the saturation magnetic field for the parameter range $-0.4 \lesssim J_2/J_1$ and $J_2>0$.
We determine the phase diagram for 3-dimensional model at high magnetic field by representing spin flips out of the saturated state as bosons, considering the dilute boson limit and using the Bethe-Salpeter equation to determine the first instability of the saturated paramagnet.
Close to the highly frustrated point $J_2/J_1\sim0.5$, we find that the spin-nematic state is stable even for $|J_{\text{C}}/J_1|\sim 1$.
For larger values of $J_2/J_1$, interlayer coupling favors a broad, phase-separated region.
Further increase of $|J_{\text{C}}|$ stabilizes a collinear antiferromagnet, which is selected via the order-by-disorder mechanism.
}

\kword{frustration, spin nematic}

\begin{document}
\maketitle
\begin{figure}[ht]
\begin{center}
\includegraphics[scale=0.22]{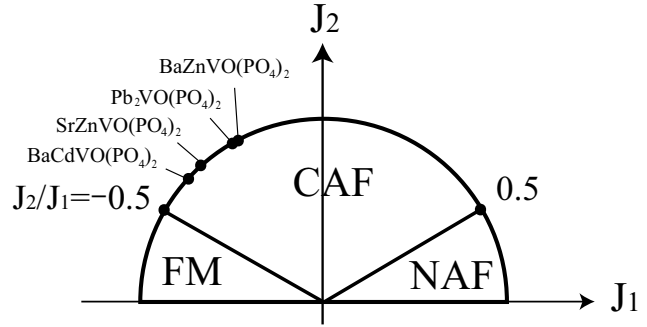}
\caption{
Classical phase diagram of the $J_1$-$J_2$ square lattice model [Eq.~\ref{HSpin}] 
for $J_2>0$ and ${\text H}=0$.
FM, NAF, and CAF stand for ferromagnetic, 
N\'{e}el antiferromagnetic, and collinear anti-ferromagnetic phases.
The spin configuration of each phase 
is shown in Fig.~\ref{Fig:config}.
Also shown are the experimentally determined exchange parameters of several materials \cite{J1J2compound_1} whose magnetic properties are well described by $H_{2d}$ [Eq.~\ref{HSpin}]:
$J_2/J_1=-0.9$ for BaCdVO(PO$_4$)$_2$; 
$J_2/J_1=-1.1$ for SrZnVO(PO$_4$)$_2$;
$J_2/J_1=-1.8$ for Pb$_2$VO(PO$_4$)$_2$;
$J_2/J_1=-1.9$ for BaZnVO(PO$_4$)$_2$.
In the $S=1/2$ quantum case, 
for $J_2/|J_1|\gtrsim 0.4$ and $J_1<0$, 
the spin nematic phase is 
theoretically expected 
slightly below the saturation phase\cite{SquareNem,comm1}
\label{Fig:phaseD}}
\end{center}
\end{figure}

{\it Introduction}-
The combination of frustration and quantum fluctuations often leads to exotic magnetic phases.
One example is the spin-nematic state, in which spin operators have zero expectation values, but components of a rank-2 tensor formed from products of spin operators have non-zero expectation values \cite{Andreev,review_nem}. 
Theoretically, the spin-nematic state has been shown to exist in various  frustrated-Heisenberg models.
One example is the frustrated spin-$1/2$ \mbox{$J_1-J_2$} model on the square lattice,
\begin{equation}
\begin{split}
H_{2d}=&\sum_{\text{n.n.}} J_{1} {\bf S}_{\boli}\cdot {\bf S}_{\bolj}
+\sum_{\text{n.n.n.}} J_{2} {\bf S}_{\boli}\cdot {\bf S}_{\boli^\prime}
+{\text H}\sum_\boll S^z_\boll\ ,
\label{HSpin}
\end{split}
\end{equation}
where `n.n. (n.n.n.)' implies (next) nearest-neighbor 
couplings in the a-b plane, 
and ${\text H}$ is an external magnetic field.
In this model there is a highly frustrated point at \mbox{$J_2/J_1=-0.5$}.
Classically, this corresponds to the phase boundary between a ferromagnetic (FM)
and a collinear anti-ferromagnetic (CAF) phase [see Fig.~\ref{Fig:phaseD}].
In the spin-1/2 model with ${\text H}=0$, 
it has been theoretically argued that
a spin-nematic state appears between the FM and CAF phases for a narrow parameter range, although the existence of the nematic phase at zero field is still under debate\cite{SquareNem,Richtersquare,Feldner,sm,Ueda_Totsuka,ShindouM}.
Close to saturation, the spin-nematic state is stable for a much larger parameter range $0.4\lesssim J_2/|J_1|$ and $J_1<0$.
This has been shown both by exact diagonalisation and by analytic calculation of the magnon binding energy in the saturated state \cite{SquareNem,Thalmeier,comm1}.
In this analytic approach, the energy of the bound magnon state is calculated exactly \cite{Matis,Bethe}, and if the energy gap to bound magnon excitations closes at a higher magnetic field than the single-magnon (spinwave) gap,  the spin-nematic state appears.

There are several compounds that approximately realize 
the square-lattice, spin-1/2 $J_1$-$J_2$ model \cite{J1J2compound_1,J1J2compound_2,J1J2compound_3,Oba-Tsujimoto}.
Materials with $J_1<0$ include 
BaCdVO(PO$_4$)$_2$, 
SrZnVO(PO$_4$)$_2$,
Pb$_2$VO(PO$_4$)$_2$,
and BaZnVO(PO$_4$)$_2$,
and their estimated exchange couplings [see Fig.~\ref{Fig:phaseD}] suggest they may host spin-nematic phases at high magnetic field \cite{J1J2compound_1}.
Recently, several techniques have been proposed to detect the spin-nematic state\cite{MSato,Andy}, and there is hope that the experimental realization of this phase could occur in the near future.

\begin{figure}[ht]
\begin{center}
\includegraphics[scale=0.34]{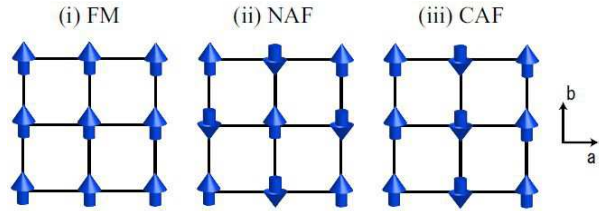}
\caption{(Color online)
Spin configurations of the classical ground states of $H_{2d}$ [Eq.~\ref{HSpin}] at ${\text H}=0$ (see Fig.~\ref{Fig:phaseD}).
\label{Fig:config}}
\end{center}
\end{figure}

In any real compound, 
there is always a finite interlayer coupling. 
This is the case for BaCdVO(PO$_4$)$_2$, 
SrZnVO(PO$_4$)$_2$, 
Pb$_2$VO(PO$_4$)$_2$
and BaZnVO(PO$_4$)$_2$.
Naively, this would tend to destabilize non-trivial quantum phases, and thus, in order to guide the experimental search for the spin-nematic state, it is important to study the effect of interlayer coupling.
The role of interlayer coupling on $H_{2d}$ [Eq.~\ref{HSpin}] has been studied in the classical CAF, 
and N\'{e}el antiferromagnetic (NAF) phases, as well as in the quantum disordered phase near the CAF/NAF boundary \cite{Thalmeier,Schmalfus,Holt,Nunes,Majander}.
However, to our knowledge, it has not been studied in the spin-nematic phase.
This is unlike the case of quasi-1D $J_1$-$J_2$ chains, where the stability of the spin-nematic state to interlayer coupling has been studied extensively \cite{Kuzian,Zhito_Tsune,Nishimoto,Syromyatnikov,Sato2,HTUandKT,Starykh2,HTU_KT_3}.

\begin{figure}[ht]
\begin{center}
\includegraphics[scale=0.38]{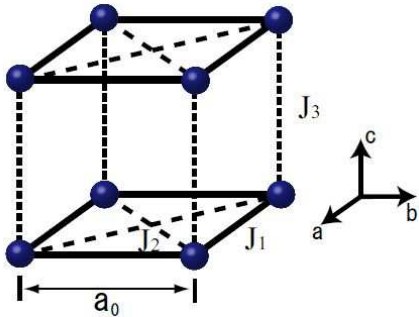}
\caption{(Color online)
Three-dimensional stacked-square (cubic) lattice. 
Filled spheres denote spins connected by Heisenberg
exchange interactions. 
$J_1$ ($J_2$) describes the (next) nearest-neighbor exchange interaction in the a-b plane.
The interlayer coupling is given by $J_{\text{C}}$. 
We set the lattice constant $a_0=1$.
\label{Fig:cubic}}
\end{center}
\end{figure}

In this Letter, we study the effect of interlayer coupling 
on $H_{2d}$ [Eq.~\ref{HSpin}] close to the CAF/FM phase boundary 
in high magnetic field, fully taking into account quantum fluctuations. 
We consider a cubic lattice formed from $J_1$-$J_2$ planes with interlayer coupling $J_{\text{C}}$ (see Fig.~\ref{Fig:cubic}).
We determine the phase diagram just below the saturation field using the dilute-Bose-gas 
and Bethe-Salpeter (bound-magnon) methods \cite{Ueda_Momoi,Batyev,Nikuni-Shiba-2,HTUandKT}.
We find that the spin-nematic state is robust close to the classical CAF/FM boundary ($J_2/J_1\sim -0.5$), and is the ground state even for $|J_{\text{C}}/J_1| \sim1$.
At higher values of $J_2/J_1$, the spin-nematic state is destabilized by large interlayer coupling $|J_{\text{C}}|$, and we find a sizeable region of parameter space where a phase-separated state is expected.
For large values of $|J_{\text{C}}|$ the semiclassically expected canted-CAF phase appears.



{\it Hamiltonian-}
We study the stacked $J_1$-$J_2$ Heisenberg model on the square lattice with interlayer coupling $J_{\text{C}}$ (i.e the cubic lattice, see Fig.~\ref{Fig:cubic}),
\begin{equation}
\begin{split}
H=&\sum_{\text{n.n in a-b}} J_{1} {\bf S}_{\boli}\cdot {\bf S}_{\bolj}
+\sum_{\text{n.n.n in a-b}} J_{2} {\bf S}_{\boli}\cdot {\bf S}_{\boli^\prime}
+\sum_{\boll} J_{\text{C}} {\bf S}_{\boli}\cdot {\bf S}_{\bolj+\bole_c}\\
&+{\text H}\sum_\boll S^z_\boll\ ,
\label{HSpin}
\end{split}
\end{equation}
where `n.n. (n.n.n.) in a-b' implies (next) nearest-neighbor 
couplings in the a-b plane.

We use the hardcore-boson representation,
\begin{equation}
S^z_l=-1/2+a^\dagger_l a_l \; , \;\;
S_l^+ =a_l^\dagger \; , \; \;  S_l^- =a_l \; ,
\label{Ch1:hardcoreSpin}
\end{equation}
\begin{equation}
\begin{split}
H &= \sum_{q}(\omega (\bolq) - \mu)a^\dagger_{\bolq}a_{\bolq}
+\frac{1}{2N}  \sum_{\bolq,\bolk,\bolk^\prime}  V_{\bolq} 
a_{\bolk+\bolq}^\dagger a_{\bolk^\prime-\bolq}^\dagger
a_{\bolk}a_{\bolk^\prime},\\
&\omega(\bolq) =\epsilon(\bolq)-\epsilon_{\text{min}}\ ,\ \ \ 
\mu={\rm H}_{\text{c}}-{\rm H}\ ,\\
&{\rm H}_{\text{c}}=\epsilon({\bf 0})-\epsilon_{\text{min}}\ ,\ \ \ 
V_{\bolq} =2(\epsilon(\bolq)+U)\ ,
\end{split}
\label{Hboson}
\end{equation}
where the on-site interaction $U\rightarrow \infty$ and,
\begin{equation}
\begin{split}
\epsilon({\bf q}) &= 
J_1 (\cos q_a + \cos q_b) 
+ J_2 (\cos (q_a+q_b) + \cos (q_a-q_b))\\
&+J_{\text{C}}\cos q_c\ ,
\end{split}
\end{equation}
with $\epsilon_{min}$ the minimum of $\epsilon(\bolq)$:

(i) For $-2\leq J_1/J_2\leq2$ and $J_2>0$: 
$\epsilon_{min}=\epsilon(\bolQ_{\pm}^{(f,a)})=-2J_2-|J_{\text{C}}|$, 
where the labels (f) and (a) are respectively chosen 
for $J_{\text{C}}<0$ and $J_{\text{C}}>0$. 
$\bolQ_{+}^{(f)}=(\pi,0,0)$, $\bolQ_{-}^{(f)}=(0,\pi,0)$, 
$\bolQ_{+}^{(a)}=(\pi,0,\pi)$ and $\bolQ_{-}^{(a)}=(0,\pi,\pi)$.

(ii) For $J_1/J_2\leq-2 $ and $J_2>0$: 
$\epsilon_{min}=\epsilon(\bolQ_f^{(f,a)})=2J_1+2J_2-|J_{\text{C}}|$,
where $\bolQ_f^{(f)}=(0,0,0)$ and 
$\bolQ_f^{(a)}=(0,0,\pi)$. \\
Here ${\text H}_c$ is the saturation field. 
If the field is reduced below $\text{H}_c$, 
the magnon gap closes ($\mu>0$),
and magnon-Bose-Einstein condensation may occur.



{\it GL Analysis-}
We focus here on the case $-2\leq J_1/J_2\leq 2$ and $J_{2,3}>0$.
An equivalent analysis can be made for $J_{\text{C}}<0$. 
Slightly below the saturation field, and for $\mu>0$,
Bose-Einstein condensation of magnons may occur at two momenta,
\begin{align}
\VEV{a_{\bolQ_+^{(a)}}}&=\sqrt{N\rho_{\bolQ_+}}\exp(i\theta_{\bolQ_+}),\\
\VEV{a_{\bolQ_-^{(a)}}}&=\sqrt{N\rho_{\bolQ_-}}\exp(i\theta_{\bolQ_-}).
\end{align}
The induced spin-ordered phase is characterized by the wave vectors 
$\bolQ_+^{(a)}$ and/or $\bolQ_-^{(a)}$.

In the dilute limit, the energy density $E/N$ is expanded in the density
$\rho_{{\bf Q}_{\pm}}$.
Retaining terms up to quadratic order gives,
\begin{align}
%
\frac{E}{N} =&\frac{\Gamma_1}{2}
\left(\rho_{\bolQ_+}^2+\rho_{\bolQ_-}^2\right)
+[ \Gamma_2 +\Gamma_3\cos 2(\theta_{\bolQ_+}-\theta_{\bolQ_-}) ]
\rho_{\bolQ_+} \rho_{\bolQ_-} \nonumber\\
&- \mu(\rho_{\bolQ_+}+\rho_{\bolQ_-}).
\label{Ch2:EffPotential}
\end{align}
Here we introduced the renormalized interactions $\Gamma_1$, which acts between bosons of the same species, $\Gamma_2$ which acts between different species,
and $\Gamma_3$, which describes umklapp scattering.

\begin{figure}[ht]
\begin{center}
\includegraphics[scale=0.5]{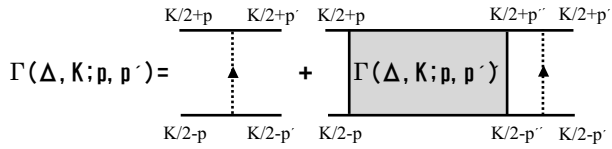}
\caption{(Color online)
Ladder diagram for the scattering amplitude $\Gamma$ [Eq.~\ref{Ch2:laddereq}].
\label{Fig:ladder}}
\end{center}
\end{figure}

$\Gamma_{1,2,3}$ are determined from the scattering amplitude 
shown in Fig.~\ref{Fig:ladder},
\begin{equation}
\begin{split}
&\Gamma(\Delta,\bolK;\bolp,\bolp^\prime)=V(\bolp^\prime-\bolp)+V(-\bolp^\prime-\bolp)\\
&-\frac{1}{2}\int \frac{d^3 p^{\prime\prime}}{(2\pi)^3}
\frac{\Gamma(\Delta,\bolK;\bolp,\bolp^{\prime\prime})[V(\bolp^\prime-\bolp^{\prime\prime})
+V(-\bolp^\prime-\bolp^{\prime\prime})]}
{\omega(\bolK/2+\bolp^{\prime\prime})+\omega(\bolK/2-\bolp^{\prime\prime})+\Delta-i0^+},
\end{split}
\label{Ch2:laddereq}
\end{equation}
where the integral is taken over the region $p^{\prime\prime}_{x,y,z}\in (0,2\pi)$. 
$\bolK$ is the center-of-mass momentum
of the two magnons and $\Delta$ is the binding energy.
We solve this integral exactly \cite{Ueda_Momoi,Batyev,Nikuni-Shiba-2,HTUandKT}.
As a result, we obtain
\begin{equation}
\begin{split}
\Gamma_1 &=\Gamma(0,2\bolQ_+^{(a)};0,0)/2\ ,\\
\Gamma_2 &=\Gamma(0,\bolq_1;\bolq_2,\bolq_2)\ ,\\
\Gamma_3&=\Gamma(0,2\bolQ_+^{(a)};0,\bolq_0)/2\ ,
\end{split}
\label{Gam3}
\end{equation}
where $\bolq_0=(\pi,\pi,0)$, $\bolq_1=(\pi,\pi,2\pi)$, and $\bolq_2=(\pi/2,-\pi/2,0)$.

The values of $\Gamma_{1,2,3}$ determine the nature of the emergent phase for $\mu>0$. 
When $\Gamma_1<\Gamma_2-|\Gamma_3|$ and $\Gamma_1>0$, 
$\rho_{Q_+}=\rho=\frac{\mu}{\Gamma_1}$ and $\rho_{Q_-}=0$ 
(or vice versa).
When the magnon at the wavevector $\bolQ_+^{(a)}$ condenses as,
\begin{equation}
\VEV{a_l}=\sqrt{\rho}\exp[i({\bf Q_+^{(a)}}{\cdot}{\bf R}_l
+\theta_{\bolQ_+})],
\end{equation}
the spin-expectation values are given by,
\begin{equation}
\begin{split}
& \VEV{S_l^z}=-\frac{1}{2}+\rho\ ,\\
& \VEV{S_l^x} = \sqrt{\rho}\cos ({\bf Q_+^{(a)}}{\cdot}{\bf R}_l 
+ \theta_{\bolQ_+})\ ,\\
& \VEV{S_l^y} = -\sqrt{\rho}
\sin ({\bf Q_+^{(a)}}{\cdot}{\bf R}_l + \theta_{\bolQ_+}).\\
\end{split}
\end{equation}
This describes the canted-CAF phase, in agreement with predictions from large-$S$ spin-wave theory via the order-by-disorder mechanism \cite{Shender,Henley}.

If $\Gamma_1>\Gamma_2-|\Gamma_3|$, $\Gamma_1>0$ and 
$\Gamma_1+\Gamma_2-|\Gamma_3|>0$, 
$\rho_{Q_+}=\rho_{Q_-}=\rho^\prime
=\frac{\mu}{\Gamma_1+\Gamma_2-|\Gamma_3|}$.
In this case, 
we expect a nontrivial multiple-Q (double-Q) 
phase, which is also observed 
in several other models\cite{HTUandKT,Nikuni-Shiba-2,Veillette,Kamiya,Marmorini,Starykh,Yamanaka,Martin,Akagi,Hayami}. 
However, these values of $\Gamma_{1,2,3}$ are not realised in $H$ [Eq.~\ref{HSpin}].

When $\Gamma_1<0$ or $\Gamma_1+\Gamma_2-|\Gamma_3|<0$, 
the dilutely-condensed phase is unstable, and a jump in 
the magnetization curve (phase separation) is expected at $\mu<0$.\cite{Ueda_Momoi}  
This follows from the divergence of $E/N$ [Eq.~\ref{Ch2:EffPotential}], which is in turn due to the lack of higher-order interaction terms. 
For example, if $\Gamma_1<0$, it can be seen that $E/N\to -\infty$ if $\rho_\bolQ \to \infty$.

{\it Bound Magnon-}
We have discussed the magnetic phases induced by {\it single} magnon condensation
just below the saturation field. 
The other possibility is that magnons form stable-bound states, and the gap to the bound magnon closes 
at higher field than that of the single magnon.
As a consequence, the bound magnon can condense, leading to spin-nematic state with a director order parameter perpendicular to the field. 
The order parameter is given by
$\VEV{S^\pm}_i=0$, 
$\VEV{S^+_i S^+_j}\neq 0$.

The binding energy and the wavefunction of the two-magnon bound state can
be understood from the scattering amplitude $\Gamma$. 
The divergence of $\Gamma$ implies a stable bound state
with binding energy $\Delta_B(\bolK)$.
If the largest binding energy has $\Delta_{\text{min}}>0$, 
the bound state will condense when 
$\text{H}<\text{H}_{c2}=\text{H}_c+\Delta_{\text{min}}/2$. 
The wavefunction of the bound state follows from the residue of $\Gamma$.\cite{Bethe}

\begin{figure}[ht]
\begin{center}
\includegraphics[scale=0.77]{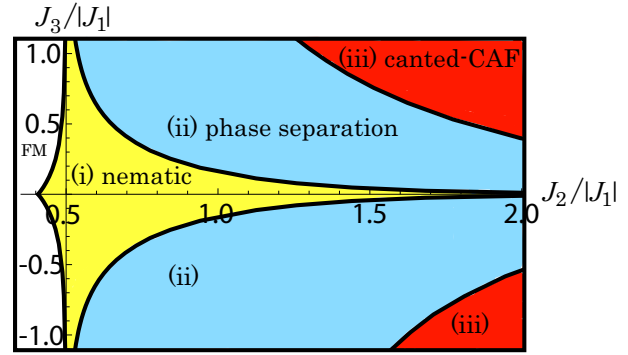}
\caption{(Color online)
Phase diagram of $H$ [Eq.~\ref{HSpin}] slightly below the saturation field and with $J_1<0$. 
The phases are:
(i) spin nematic; 
(ii) phase separation (1st-order phase transition); 
(iii) canted-CAF phase (expected from the large-S expansion). 
In the unlabeled white region, 
the trivial FM (antiferromagnetic phase along $c$-axis) is expected 
for $J_{\text{C}}<0$ ($J_{\text{C}}>0$).
\label{Fig:phase1}}
\end{center}
\end{figure}

\begin{figure}[ht]
\begin{center}
\includegraphics[scale=0.85]{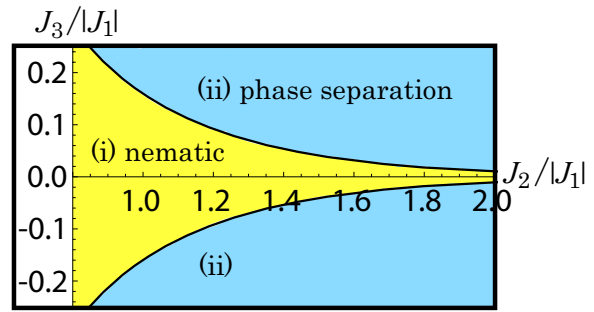}
\caption{(Color online)
Expanded view of Fig.~\ref{Fig:phase1} at small $J_{\text{C}}/|J_1|$.
\label{Fig:phase2}}
\end{center}
\end{figure}

{\it Phase Diagram-}
By calculating $\Gamma_{1,2,3}$ numerically, 
we obtain the phase diagram slightly below the saturation field, and this is shown  
in Figs.~\ref{Fig:phase1},\ref{Fig:phase2}.
In the yellow region (i), the bound magnon
is the leading instability of the fully polarized phase. 
\cite{commentNem}
Near the classical CAF/FM phase boundary ($J_2/|J_1|\sim 0.5$), 
the spin-nematic phase exists even at $|J_{\text{C}}/J_1|\sim 1$.

In the blue region (ii), $\Gamma_1<0$, and a phase separation is expected.
In consequence, there is a magnetization jump when the magnetic field is lowered through the saturation value.
It is beyond the scope of this Letter to predict 
which phase occurs below saturation, since the first-order-phase transition
introduces a finite density of magnons, and the dilute Bose gas approximation breaks down. 

For $\Gamma_1<\Gamma_2-|\Gamma_3|$, a naive approach that neglects the effect of finite density suggests 
the 1st-order phase transition to canted CAF with 
an associated jump in the magnetization.
However, we cannot exclude the possibility that a spin-nematic phase or a double-Q phase is stabilized by interaction effects.
On the boundary of the (i) nematic and (ii) phase separation regions, the s-wave scattering amplitude $\Gamma_1$ diverges, and, close to this boundary, the Efimov effect is expected \cite{Nishida}.

In the (iii) red region, single magnons condense and form 
a canted-CAF phase. 
The phase (i), (ii) and (iii) span 
the entire region 
where the canted-CAF phase is expected semiclassically ($-2<J_1/J_2<2$ and $J_2>0$, see Fig.~\ref{Fig:phaseD}).
For $0<J_1/J_2<2$ and $J_1>0$ the first instability of the saturated paramagnet is always to the canted-CAF phase.
This is true even in the highly frustrated region $J_1/J_2\sim 2$ (classical CAF-NAF phase boundary in Fig.~\ref{Fig:phaseD}).


{\it Conclusion-}
We have studied the effect of interlayer coupling, $J_{\text{C}}$,
on the magnetic phase diagram of the $S=1/2$ stacked-square-lattice 
$J_1$-$J_2$ model under high external field, using the dilute Bose-gas technique.\cite{Batyev,Bethe}
The main result, shown in
Figs.~\ref{Fig:phase1},\ref{Fig:phase2},
is the phase diagram just below the saturation field.
While semi-classical theory always predicts a canted CAF phase, a full quantum treatment reveals the presence of spin-nematic and phase separated regions.
The spin-nematic state, which has previously been shown to exist in the pure 2D model ($J_{\text{C}}=0$) \cite{SquareNem},
is robust against the addition of interlayer coupling in the vicinity of the FM/CAF phase boundary ($J_2/J_1\sim-0.5$).
For larger values of $J_2/|J_1|$ a broad phase-separation region is stabilized by the addition of $J_{\text{C}}$ coupling. 
Here a magnetization jump is expected as field is lowered through the saturation value, and, below this jump, the canted CAF-phase is expected to appear, although interactions may favor a spin nematic or double-Q phase.
On the boundary between the spin-nematic phase and the phase separated region, 
the s-wave scattering amplitude $\Gamma_1$ diverges and 
the Efimov effect is expected.\cite{Nishida}
The final conclusion is that in a quasi-2D, $J_1$-$J_2$ compound with $J_1<0$ and $J_2/|J_1|\gtrsim 0.5$, close to the saturation magnetic field the spin nematic state is remarkably robust against interlayer coupling.

The author thanks N.~Shannon and A.~Smerald
for useful discussions and careful readings of this Letter. 
In addition, the author appreciates the helpful English corrections 
of this Letter by A.~Smerald.
The author is also grateful to
supports by Okinawa Institute of Science and Technology, 
and JSPS KAKENHI Grant No. 26800209.


\begin{thebibliography}{9}
\bibitem{comm1}
In Ref.~\citen{SquareNem}, 
the parameter range of $J_2/J_1$ where the spin nematic phase
appears slightly below the saturation field 
is not declared explicitly.
The concrete parameter is reproduced in Ref.~\citen{Ueda_Momoi} 
and in this letter.
\bibitem{commentNem}
We do not exclude the possibility of 
the 1st-order phase transition at the saturation field 
even in the yellow region (i). 
It happens 
if the not-evaluated interaction between the bound magnon is negative. 
Actually, for $J_2/|J_1|\lesssim 0.5$ at $J_{\text{C}}=0$, 
the 36-sites exact diagonalization suggests 
the 1st-order phase transition to the nematic phase at the saturation 
field\cite{SquareNem}.
Hence, in our stacked-lattice case for $J_{\text{C}}\neq 0$, 
we may expect the 1st-order phase transition to the nematic phase 
for $J_1/|J_2|\lesssim 0.5$
in the yellow region (i).
\end{thebibliography}
\end{document}